\documentclass[preprint,5p,twocolumn]{elsarticle}
\usepackage{hyperref}
\usepackage{lineno}
\usepackage{upgreek}
\usepackage{float}
\usepackage{bm}% bold math
\usepackage{xcolor}
\usepackage{ulem}
\usepackage{amsmath} 
\usepackage{graphicx}
\usepackage[english]{babel}  
\journal{arXiv}
\begin{document}
%\begin{linenumbers}
 
\title{Deep diving into the comparative study of Choline dynamics using molecular dynamics simulation and neutron scattering technique}
\author[fn1]{Debsindhu Bhowmik\corref{mycorrespondingauthor1}}
\cortext[mycorrespondingauthor1]{Corresponding author}
\ead{bhowmikd@ornl.gov}   
\address[fn1]{Computational Science and Engineering Division, Oak Ridge National Laboratory, Oak Ridge,  Tennessee 37831, USA.}   
\date{\today}

\begin{abstract}

We present here the comparative study between the dynamics Choline and Tetra-methyl ammonium bromide. This is well known that deficiency in Choline would cause many severe diseases. No wonder why Choline is crucial component for our nutrients and dietary requirements. We present here a comprehensive study using all-atom molecular dynamics simulation combined with neutron scattering technique for solute behavior in aqueous solution. The solvent behavior is discussed in the follow up work.   

\end{abstract}

\maketitle

%\section*{Notice of Copyright}
%{Manuscript has been authored by UT-Battelle, LLC under Contract No. DE-AC05-00OR22725 with the U.S. Department of Energy. The United States Government retains and the publisher, by accepting the article for publication, acknowledges that the United States Government retains a non-exclusive, paid-up, irrevocable, worldwide license to publish or reproduce the published form of this manuscript, or allow others to do so, for United States Government purposes. The Department of Energy will provide public access to these results of federally sponsored research in accordance with the DOE Public Access Plan (http://energy.gov/downloads/doe-public-access-plan)}

\section{\label{Intro} Introduction}    

\begin{figure}[!htbp]
  \begin{center}
    \includegraphics[width=0.45\textwidth,angle=0]{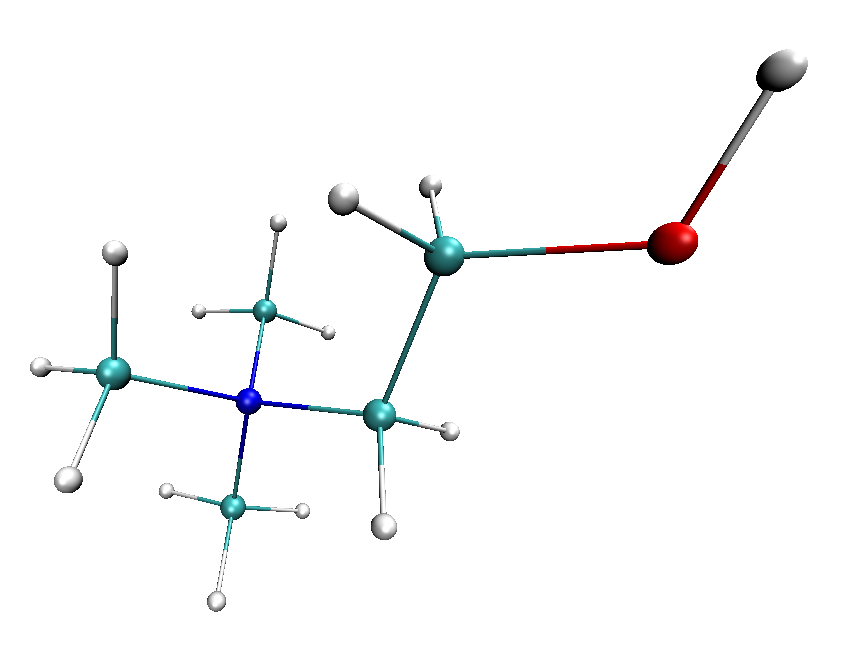} 
  \end{center}
  \bf\caption {Choline$^{+}$ cation. blue - nitrogen, green - carbon, white - hydrogen and red - oxygen atom.}
  \label{fig:choline}
\end{figure}

Choline (figure~\ref{fig:choline}) is an essential nutrient that is soluble in water and prominent component of our dietary requirement for methyl groups. This particular substance belongs to the tetra-alkyl ammonium (TAA) family i.e. contains \texttt{quaternary amine} that is crucial for several key biological processes like metabolism, lipid transport, signaling functions of cell membranes etc. These are also important during pregnancy and development of fetus. Deficiency in Choline could cause many diseases related to liver including neurological disorders. Free Choline, phosphocholine and glycerophosphocholine are the primary structures of Choline found in human milk.                

In this work we talk about the comparison between TMABr (the smallest in the TAA family~\cite{Bhowmik6}) with CholineBr (one may consider close derivative of TMABr with one additional -CH$_2$(OH) group attached to one arm of TMABr) in their aqueous solution.   

We will principally discuss about different kinds of dynamics namely coherent, incoherent~\cite{Bhowmik6} translational of both the cation and the anion (that is Bromide in this case). Further the solvent dynamics including the separation of translational and rotational part of the hydration water molecules around the molecule will be presented.        

The study on the dynamics will be performed by a combination of Time-of-Flight (TOF), Neutron Spin Echo (NSE) and Molecular Dynamics (MD) simulation methods~\cite{Bhowmik, Bhowmik1, Bhowmik2, Bhowmik3, Bhowmik4, Bhowmik5, Bhowmik6, Bhowmik7, Bhowmik8, Bhowmik9, Bhowmik10, Bhowmik11, Bhowmik12, Bhowmik13, Bhowmik14, Bhowmik15}. We take advantage of the above mentioned techniques to study separately the coherent and incoherent signal dynamics and will find out if the two approaches (coherent and incoherent signal dynamics) lead to two different results in comparison to with MD simulation.    

\section{\label{methods} Materials and Methods}\ 
\subsection{\label{simudetails} Simulation Details}\ 
\subsubsection{parameter models}    
The details of the MD simulation is mentioned in the referred thesis~\cite{Bhowmik6}. Quoting the thesis classical molecular dynamics (MD) simulations (using DL POLY 2.18~\cite{Smith07}) is performed on aqueous solutions of TMABr and CholineBr. An all atom (explicit N, C, H atoms and also O for Choline), flexible (bond stretch, bond bending, dihedral interaction), non-polarizable model is taken for the TMA$^{+}$ or Choline$^+$ ion. Individual atomic charges within this ion are determined by the Hartree-Fock method (for nonpolarizable force fields), followed by modification using Antechamber (AMBER routine)~\cite{Heyda10} and other interaction parameters are taken from the Generalized Amber Force Field (GAFF)~\cite{AMBER_10}. The cationic atom charges and force field parameters for TMA$^+$ and Choline$^+$ are summarized in table~\ref{tab:TMA_charge},~\ref{tab:choline_charge},~\ref{tab:TMA_FF} and~\ref{tab:choline_FF}. The sodium, bromide charges or force-field parameters are taken from earlier literatures~\cite{Koneshan98_102}~\cite{Horinek09}~\cite{Lee96} ~\cite{Joung08}~\cite{Markovich96_7}. Rigid SPC/E model (O-H bond of 1.0~\AA~\ with H-O-H angle of 109$^\circ$ and charges for hydrogens and oxygens with +0.424e and -0.848e respectively) is used for water~\cite{Berendsen87_91}. Choosing one from a large number of existing water models, is a difficult task. We decide to continue with the SPC/E model because it reproduces well both the structural and dynamic properties of bulk water over a broad range of temperatures and pressures~\cite{Brodholt93}. This is an 'extended' version of SPC model~\cite{Guillot02} where additionally an 'self-polarisation' energy correction is imposed. The non-bonding interactions in the system are described via the Coulombic and Lennard-Jones (L-J) potentials, with the use of Lorentz-Berthelot mixing rules for the L-J parameters.    

\subsubsection{simulation}    
\begin{figure}[!htbp]
  \begin{center}
    \includegraphics[width=0.45\textwidth,angle=0]{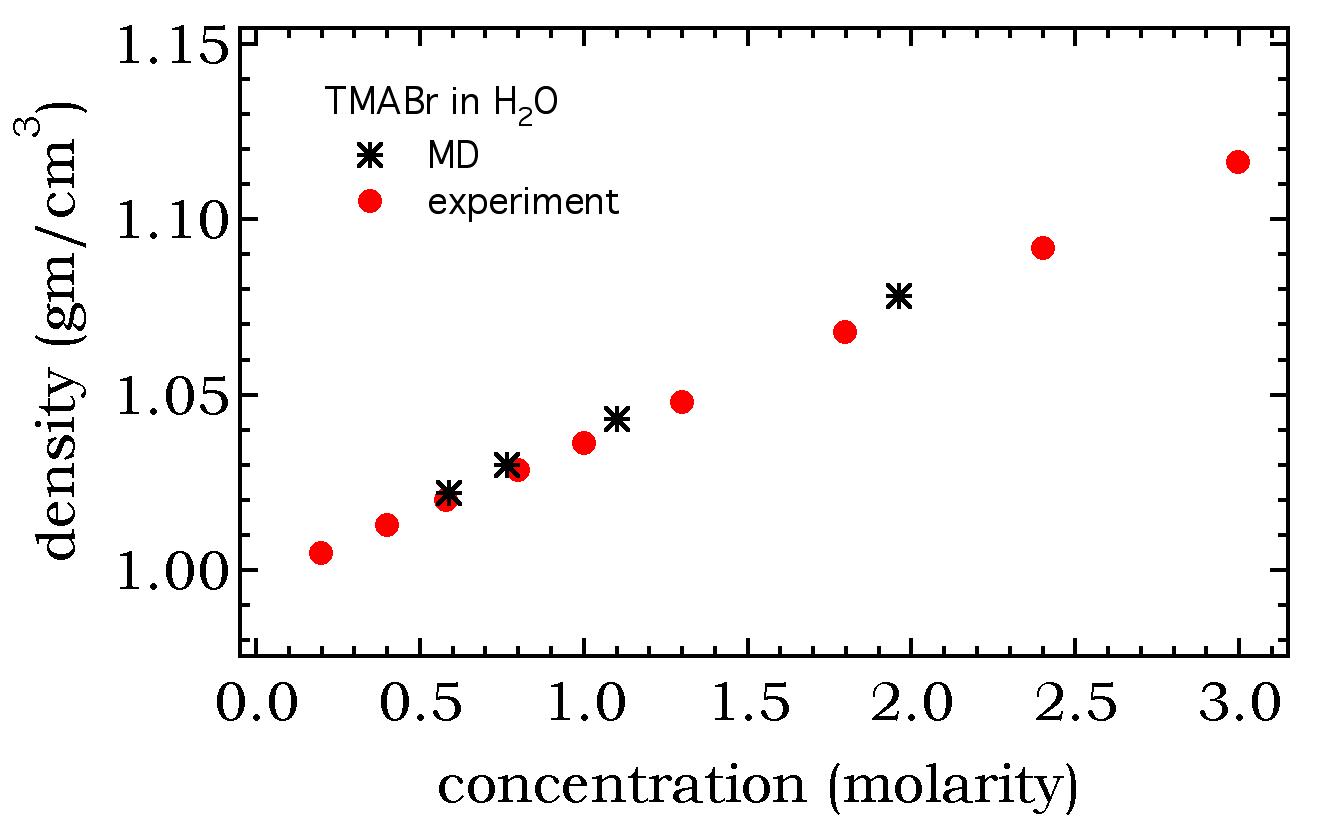} 
  \end{center}
   \bf\caption {Comparison of density of aqueous TMABr solution extracted from MD simulation and experiment~\cite{Buchner02}.}
  \label{fig:TMABr_density}
\end{figure}

\begin{figure}[!htbp]
  \begin{center}
    \includegraphics[width=0.45\textwidth,angle=0]{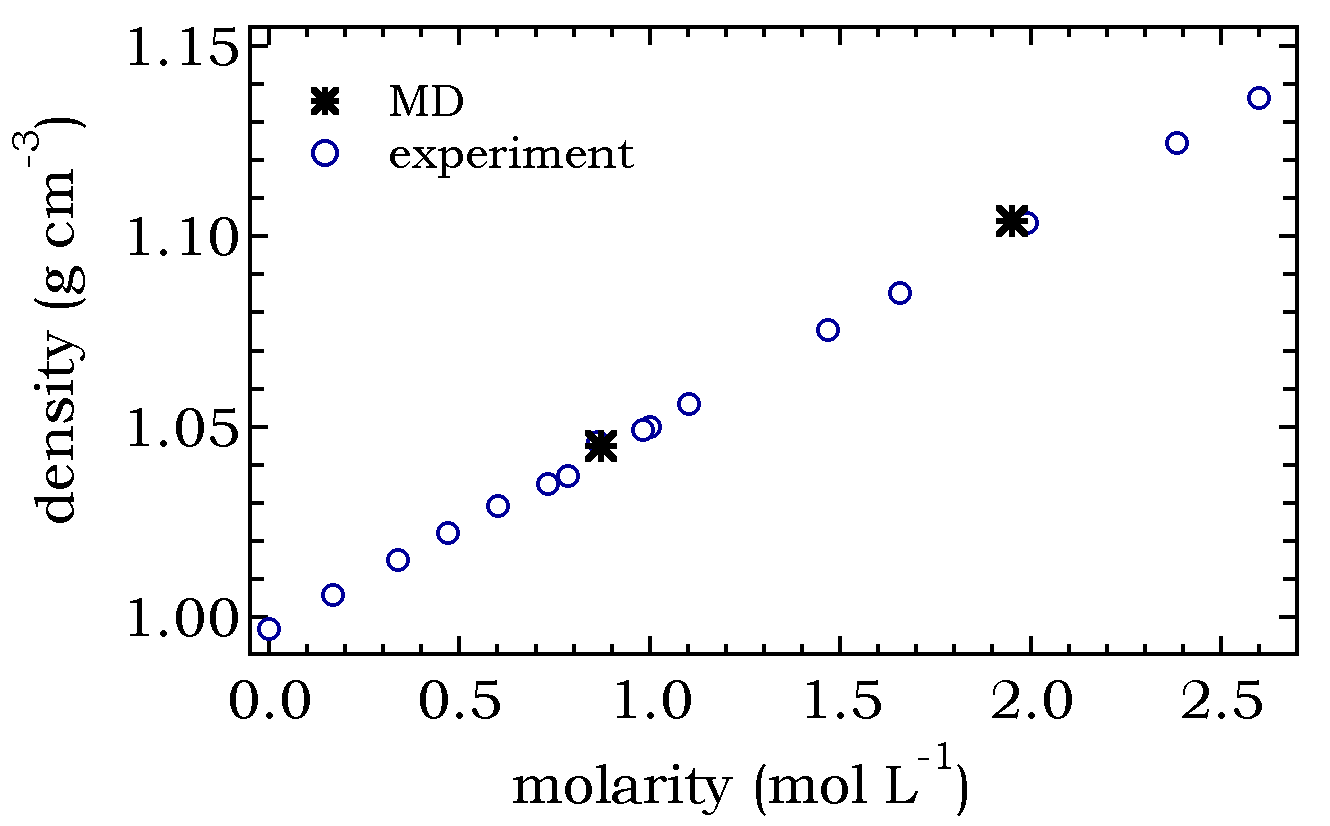} 
  \end{center}
   \bf\caption {Comparison of density of aqueous CholineBr solution extracted from MD simulation and experiment.}
  \label{fig:choline_density}
\end{figure}

Before starting any simulation it is necessary to construct the simulation box consists of all the atoms. This is done like following. First the solute ions (TAA or Choline cations) are constructed with correct geometry (in accordance to the experimentally derived bond length, valence and dihedral angles). Next a cubic simulation box is formed with a volume similar to the combined volume of desired number of solute cations, anions and solvent molecules. This simulation box is then filled by the solvent molecules (randomly oriented), leaving empty space for inserting solutes. At this time one must be careful about the fact that the initial simulation box should not very far from the equilibrium condition. Once these formalisms are completed, the MD simulation is performed in NPT ensemble and is allowed to run until the potential and kinetic energies are stable, temperature and pressure becomes constant and the system density agrees with experimental value i.e until it is equilibrated. Three dimensional periodic boundary conditions are used, a cutoff radius for short-range interactions is half the box-size, long-range part of the electrostatic interaction is evaluated using the 3D Ewald sum, SHAKE algorithm is used for rigid SPC/E water molecules. The initial configuration is equilibrated in NPT and NVT ensembles (P=1atm, T=298K), prior to a production run in the NVE ensemble of 3.4ns with a timestep of 1fs. Individual atomic trajectories are saved every 0.1ps, producing 34$\times$10$^{3}$ frames in total. Trajectories are then analysed using nMoldyn~\cite{nMOLDYN}. As an initial check of the interaction potentials, solutions of different ion concentrations are simulated and the predicted density reproduces well experimental data (difference is $<$0.2\%). In figure~\ref{fig:TMABr_density} and~\ref{fig:choline_density} a comparison between experimental and simulated densities is shown for aqueous TBABr, TMABr and CholineBr solution. 

\begin{table}[!htbp]    
\scriptsize      
\begin{tabular}{|l|ccc|}
	\hline
part of molecule & atom charge (e) & from bold part & \\
 & N & C & H \\
	\hline	
-\bf{N}- & 0.096521 & & \\
-N-$\bf{CH_{3}}$ & & -0.165381 & 0.130417 \\
	\hline
\end{tabular}
 \bf\caption {TMA$^+$ atomic charge distribution.}
 \label{tab:TMA_charge}
\end{table}

\begin{table}[!htbp]    
\scriptsize    
\begin{tabular}{|l|p{0.8cm}p{0.8cm}p{0.8cm}p{0.8cm}|}
	\hline
part of molecule & charge & (e) & & \\
 & N & C & H & O \\
	\hline	
-\bf{N}- & 0.0222 & & & \\
-N-$\bf{CH_{3}}$ & & -0.142 & 0.124 & \\
-N-$\bf{CH_{2}}$- & & -0.050 & 0.134 & \\
-N-$CH_{2}-\bf{CH_{2}}$- & & -0.183 & 0.043 & \\
-N-$CH_{2}-CH_{2}-\bf{OH}$- & & & 0.473 & -0.669\\
	\hline
\end{tabular}
 \bf\caption {Choline$^+$ atomic charge distribution.}
 \label{tab:choline_charge}
\end{table} 

\begin{table}[!htbp]    
\begin{tabular}{|c|cc|}
	\hline
bond elongation & energy & length \\
harmonic parameters & (kcal/mol/\AA$^{2}$) & (\AA) \\
	\hline	
C-H$_{N}$ & 240 & 1.090 \\
C-N & 367 & 1.471 \\
	\hline
\end{tabular}
\begin{tabular}{|c|cc|}
	\hline
bond bending & energy & angle \\
harmonic parameters & (kcal/mol/rad$^{2}$) & (degree) \\
	\hline
H$_{N}$-C-H$_{N}$ & 35 & 109.5\\
H$_{N}$-C-N & 50 & 109.5 \\
C-N-C & 50 & 109.5 \\
	\hline
\end{tabular}
\begin{tabular}{|c|cc|}
	\hline
dihedral interaction & energy & angle \\
 & (kcal/mol) & (degree) \\
	\hline
X-C-N-X & 0.15 & 0.0 \\
	\hline
\end{tabular}
\begin{tabular}{|c|cc|}
	\hline
L-J & $\epsilon$ & $\sigma$ \\
parameters & kcal/mol & \AA \\
	\hline
H$_{N}$ & 0.0157 & 1.100\\
C & 0.1094 & 1.900 \\
N & 0.1700 & 1.8240 \\
	\hline
\end{tabular}
\bf\caption {TMA$^+$ Force Field.} {Force field parameters for TMA$^+$ atoms are shown (H$_{N}$ represents the hydrogens of the carbon attached to the central N).}
\label{tab:TMA_FF}
\end{table}

\begin{table}[!htbp]    
\begin{tabular}{|c|cc|}
	\hline
bond elongation & energy & length \\
harmonic parameters & (kcal/mol/\AA$^{2}$) & (\AA) \\
	\hline	
C-H$_{N}$ & 240 & 1.090 \\
C-H$_{C}$ & 340 & 1.090 \\
C-C & 310 & 1.526 \\
C-N & 367 & 1.471 \\
C-O$_{OH}$ & 320 & 1.410 \\
C-H$_{OH}$ & 553 & 0.960 \\
	\hline
\end{tabular}
\begin{tabular}{|c|cc|}
	\hline
bond bending & energy & angle \\
harmonic parameters & (kcal/mol/rad$^{2}$) & (degree) \\
	\hline
H$_{N}$-C-H$_{N}$ & 35 & 109.5\\
H$_{C}$-C-H$_{C}$ & 35 & 109.5\\
C-C-H$_{C}$ & 50 & 109.5\\
C-C-H$_{N}$ & 50 & 109.5\\
H$_{C}$-C-O$_{OH}$ & 50 & 109.5\\
C-C-N & 80 & 111.2\\
H$_{N}$-C-N & 50 & 109.5\\
C-N-C & 50 & 109.5\\
C-C-O$_{OH}$ & 50 & 109.5\\
C-O$_{OH}$-H$_{OH}$ & 35 & 109.5\\
	\hline
\end{tabular}
\begin{tabular}{|c|cc|}
	\hline
dihedral interaction & energy & angle \\
 & (kcal/mol) & (degree) \\
	\hline
X-C-N-X & 0.15 & 0.0 \\
X-C-C-X & 0.15 & 0.0 \\
X-C-O$_{OH}$-X & 0.15 & 0.0 \\
	\hline
\end{tabular}
\begin{tabular}{|c|cc|}
	\hline
L-J & $\epsilon$ & $\sigma$ \\
parameters & kcal/mol & \AA \\
	\hline
H$_{N}$ & 0.0157 & 1.387\\
H$_{N}$ & 0.0157 & 1.100\\
H$_{OH}$ & 0.0000 & 0.0000\\
O$_{OH}$ & 0.2104 & 1.7210\\
C & 0.1094 & 1.900 \\
N & 0.1700 & 1.8240 \\
	\hline
\end{tabular}
\bf\caption {Choline$^+$ Force Field.} {Force field parameters for Choline$^+$ atoms are shown (H$_{N}$ represents the hydrogens of the carbon attached to the central N).}
\label{tab:choline_FF}
\end{table}   

Note that all simulations are done with H$_2$O as the solvent. The scattering lengths of deuterium are used for the solvent H atoms in the post-simulation analysis of atomic trajectories, to yield the comparison with scattering data (measured in D$_2$O solvent). 

\subsection{\label{expdetails} Experimental Details}\   
The experimental set-up for the neutron experiment is as follows. 

\subsubsection{\label{sample_preparation} Preparation of samples}    
The hydrogenated TMABr, CholineBr, NaBr and KBr is bought from Fluka (purity $>$99\%) and preserved at a dry place far from direct sunlight. Prior to each experiment, the salts are dried under vacuum for several hours. Then they are dissolved into liquid D$_2$O (Euriso-top, 99.9\%D) or H$_2$O (distilled) with desired solute and solvent ratio. After the preparation of each sample, the salts are stored in presence of nitrogen gas. The deuteration of the solvent is important for the incoherent QENS experiment~\cite{Bhowmik6}. The absence of exchangeable Hydrogen atoms in the TAA cation is advantageous which guarantees to keep the solute and solvent character unchangeable inside the solution.     

\subsubsection{\label{experimental_setup} Setting up experiments}
For dynamic measurements, the Neutron Spin Echo (NSE) and Time of flight (TOF) techniques are used. The details are in the thesis~\cite{Bhowmik6}. The NSE experiments are performed on MUSES (LLB-Orphee, Saclay, France) spectrometer at different temperatures (298K, 316K, 336K and 348K) under controlled Helium pressure (1 atm) varying the Q vectors from 0.2 \AA$^{-1}$ to 1.6 \AA$^{-1}$. Each of the I(Q,t) is measured for 1 day. The sample holder is 1 mm thick flat quartz cell. Aluminum cells are not used because of the deformation of the sample holder while aqueous solutions are under the neutron beam for long. Carbon-glass and quartz are used for experimental resolution at low Q ($<$0.8 \AA$^{-1}$) and high Q ($>$1.3 \AA$^{-1}$) domain respectively. The correlation time is measured up to 1100 ps. TOF measurements are carried out on MIBEMOL spectrometer in LLB-Orphee with 0.2 mm thick flat quartz cell. Like NSE, all the experiments are performed under controlled temperature and pressure. Our TOF resolution is 50 $\mu$eV (HWHM) with an incident neutron beam wavelength of 6 \AA. The experimental resolution is measured by a vanadium sample. The covered $Q$ range is from 0.49 \AA$^{-1}$ to 1.97 \AA$^{-1}$. 

For each of the above mentioned measurements, we confirm no loss of sample by comparing the sample weight at the beginning and the end of experiments.    

\section{\label{results} Results and Discussion}      
\subsection{\label{ion_structure_diff-sys} Decoupling of coherent and incoherent scattering}

\begin{figure}[!htbp]
\begin{center}
\includegraphics[width=0.35\textwidth,angle=-90] {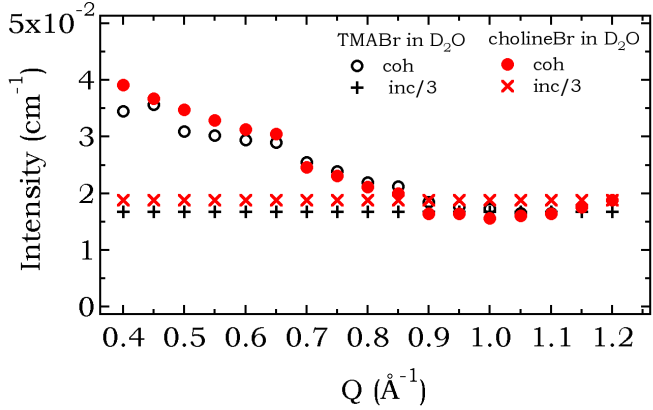}
\vspace{0.1cm}
\bf\caption{Decoupling of coherent and incoherent intensity as a function Q for aqueous TMABr and CholineBr solution with $x_m$=1:56 (extracted by MD simulation).}
\label{SQ_TMAcholine1}
\end{center}
\end{figure}

We present the application of the previously mentioned method~\ref{simudetails} to decouple the coherent and incoherent contribution for aqueous TMABr or CholineBr solution with $x_m$=1:56. Figure~\ref{SQ_TMAcholine1} shows that both the systems predict almost similar result (with CholineBr solution having slightly larger incoherent contribution because of the more hydrogen atoms than TMABr). But it should be noted that due to smaller size of the cation, the polarization is less with large uncertainty in its value and thus experimentally it is difficult to precisely separate the two contributions.   

\subsection{\label{dyn_salts_diff-conc} Translational dynamics: cation} 

\begin{figure}[!htbp]
\begin{center}
\includegraphics[width=0.45\textwidth,angle=0] {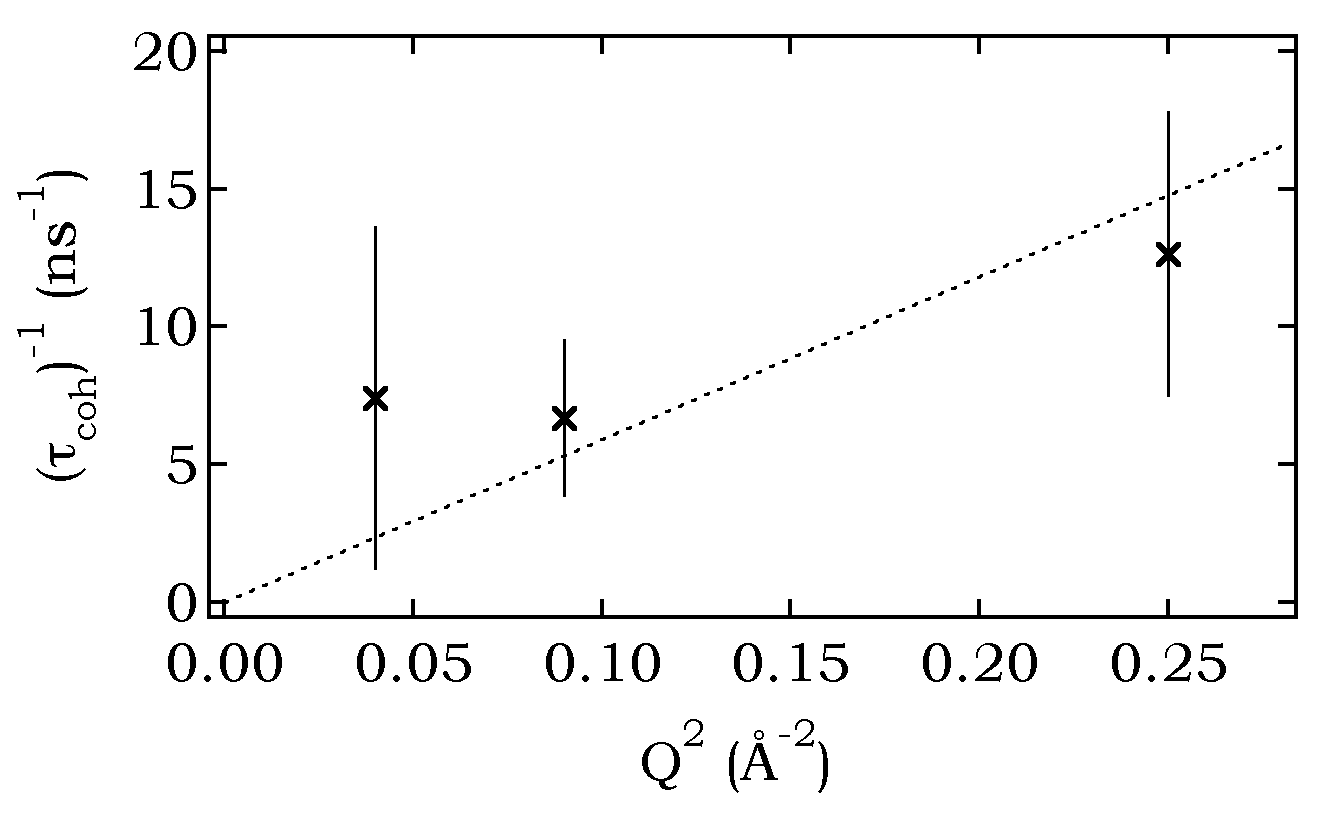}
\vspace{0.1cm}
\bf\caption{Inverse of translational relaxation time extracted from NSE coherent I(Q,t) curves is plotted as a function of $Q^2$ for aqueous TMABr solution with $x_m$=1:22. The $D_{tr}$ is extracted from a linear fit passing through origin (continuous diffusion).}
\label{coh_TMA1_NSE}
\end{center}
\end{figure}

In this section, a brief comparative study will be presented for translational dynamics in aqueous TMABr and CholineBr solution with $x_m$=1:22 and $x_m$=1:56. 

\subsubsection{NSE}      
We have carried out the NSE measurement at low Q ($<$0.6 \AA$^{-1}$) for aqueous solution of TMABr with $x_m$=1:22. In figure~\ref{coh_TMA1_NSE} we have plotted the inverse relaxation time as a function of Q$^2$ extracted from the I(Q,t) coherent analysis. This estimates $D_{tr}$=(0.72$\pm$0.10)$\times$10$^{-9}$m$^{2}$s$^{-1}$. But due to very low polarization the uncertainty in the result is very high. 

\subsubsection{ToF}    
\begin{figure}[!htbp]
\begin{center}
\includegraphics[width=0.45\textwidth,angle=0] {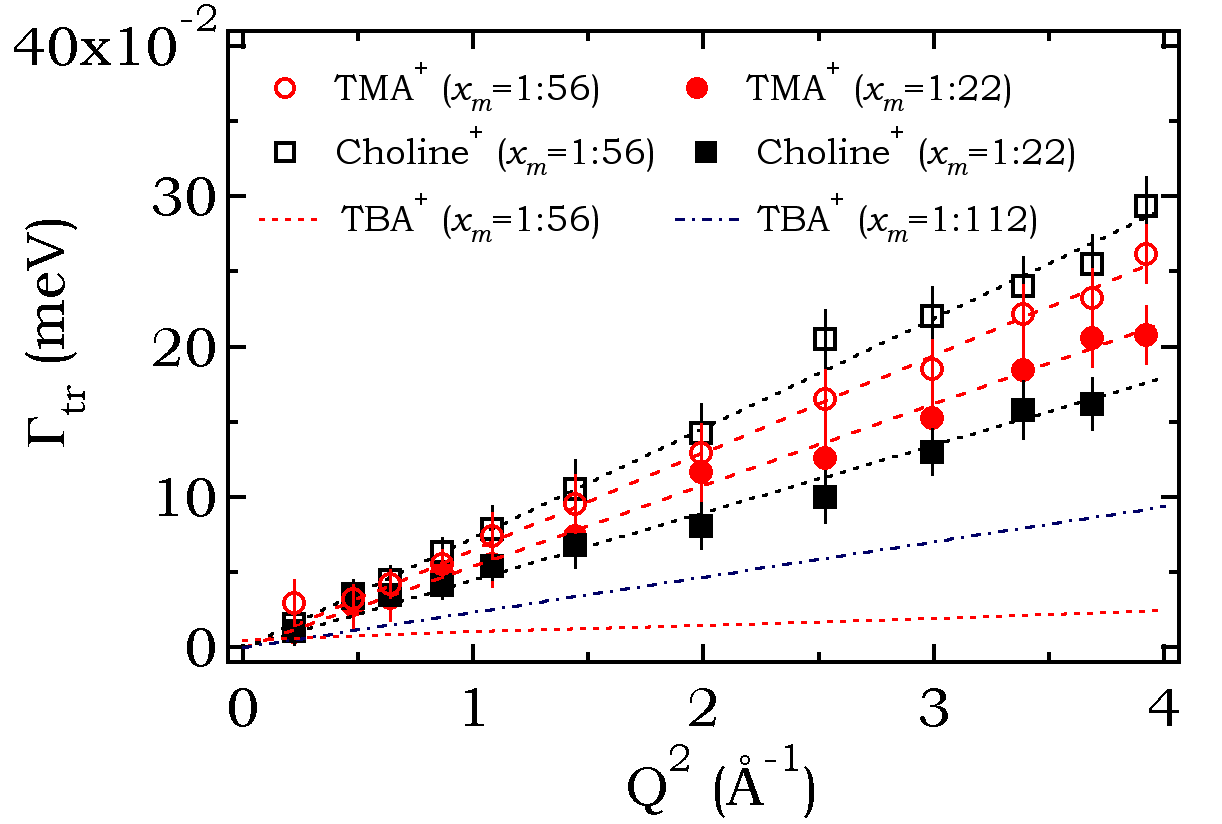}
\vspace{0.1cm}
\bf\caption{$\Gamma_{tr}$ extracted from TOF measurement for different cations at different concentrations, are plotted as a function $Q^2$. The translational diffusion coefficient $D_{tr}$ is calculated from a linear fit passing through origin (continuous diffusion).}
\label{inc_allsalt}
\end{center}
\end{figure}

Regarding this experimental difficulty in NSE coherent analysis for ions like (TMA$^+$ or Choline$^+$) we move to TOF noting the fact that the TOF could overestimates the true cation CoM translational motion~\cite{Bhowmik6}. But at the same time, we know that it can be verified by MD simulation as we did for TBA$^+$. For the TOF data fitting we have used the same model as in aqueous TBABr TOF data analysis. In figure~\ref{inc_allsalt}, we have shown the inverse translational relaxation time extracted by TOF experiment for the above mentioned systems. We have found a decrease in cationic translational diffusion coefficient with increase of concentration (as expected) and this decrease in more pronounced in case of cholineBr than TMABr. The results show that at a solution concentration with $x_m$=1:56 $D_{tr}$ for TMA and choline cation are (0.98$\pm$0.10)$\times$10$^{-9}$m$^{2}$s$^{-1}$ and (1.08$\pm$0.07)$\times$10$^{-9}$m$^{2}$s$^{-1}$ while at solution concentration with $x_m$=1:22 the values are (0.81$\pm$0.03)$\times$10$^{-9}$m$^{2}$s$^{-1}$ and (0.72$\pm$0.10)$\times$10$^{-9}$m$^{2}$s$^{-1}$ respectively.

\begin{table}   
\scriptsize    
\begin{tabular}{|c|ccc|}
	\hline
 & & $D_{tr}$ in (10$^{-9}$ m$^2$s$^{-1}$) & \\ 
 	\hline
conc. &  & individual H atom & central Nitrogen \\
        \hline	
$x_m$=1:56 & TOF & (0.98$\pm$0.10) & \\
& MSD & (0.91$\pm$0.01) & (0.75$\pm$0.01)\\
      \hline
$x_m$=1:22 & TOF & (0.81$\pm$0.07) &  \\
& MSD & (0.69$\pm$0.01) & (0.56$\pm$0.01)\\
     \hline	
\end{tabular}

 \bf\caption {Translational diffusion coefficient for TMA$^+$ calculated via coherent and incoherent analysis combing both experimental and simulation technique}
 \label{tab:TMA_diff}
\end{table}

\begin{table}   
\scriptsize    
\begin{tabular}{|c|ccc|}
	\hline
 & & $D_{tr}$ in (10$^{-9}$ m$^2$s$^{-1}$) & \\ 
 	\hline
conc. &  & individual H atom & central Nitrogen \\
        \hline	
$x_m$=1:56 & TOF & (1.08$\pm$0.03) & \\
& MSD & (0.94$\pm$0.01) & (0.67$\pm$0.01)\\
      \hline
$x_m$=1:22 & TOF & (0.72$\pm$0.10) &  \\
& MSD & (0.77$\pm$0.01) & (0.44$\pm$0.01)\\
     \hline	
\end{tabular}
 \bf\caption {Translational diffusion coefficient for Choline$^+$ calculated via coherent and incoherent analysis combing both experimental and simulation technique}
 \label{tab:choline_diff}
\end{table}

\subsubsection{Simulation with experiment}    
\begin{figure}[!htbp]
\begin{center}
\includegraphics[width=0.45\textwidth,angle=0] {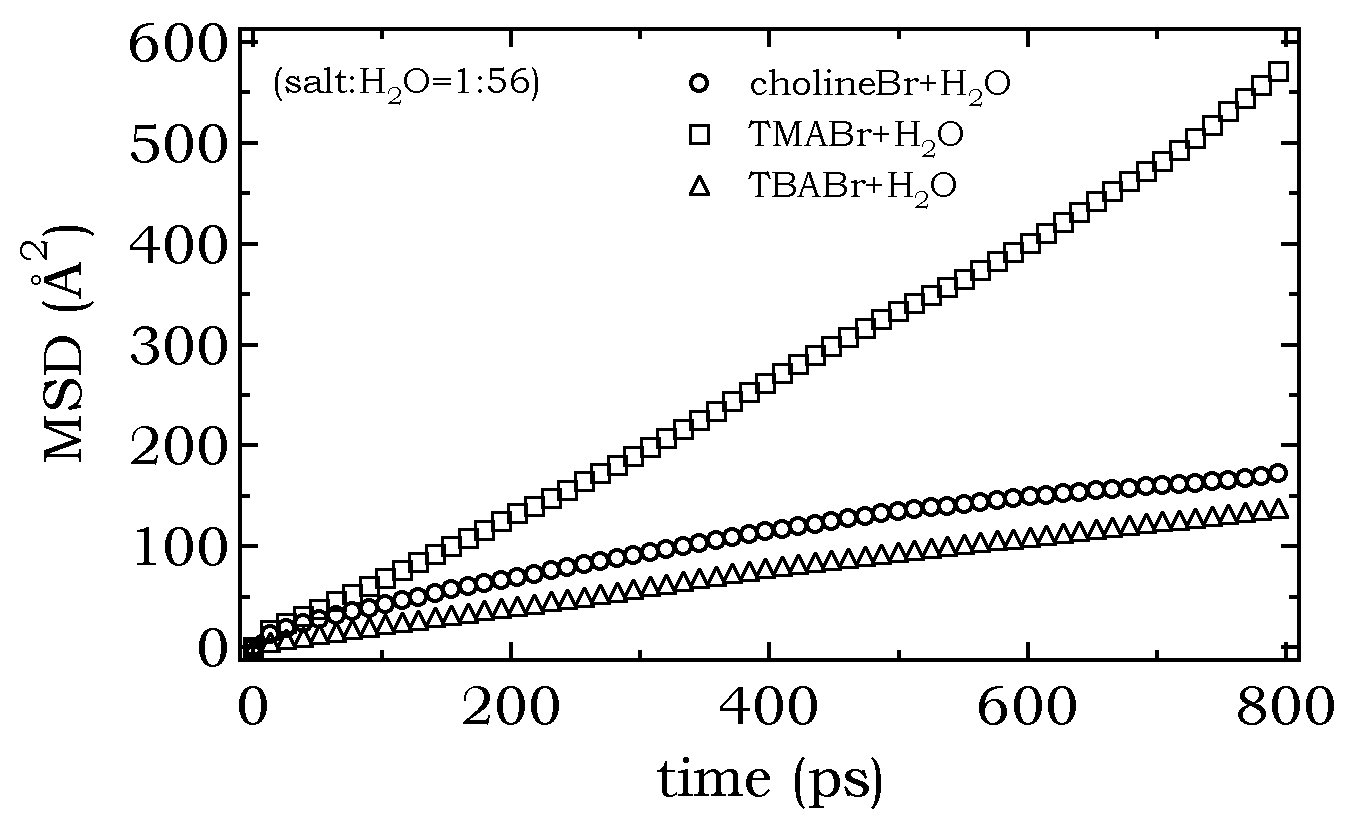}
\vspace{0.1cm}
\bf\caption{The MSD of average hydrogen atoms of TMA$^+$, Choline$^+$ and TBA$^+$ cations are plotted as a function of time for aqueous solution with $x_m$=1:56.}
\label{TBA-TMA-choline_MSD}
\end{center}
\end{figure}

Although comparing the TOF with NSE coherent result for aqueous TMABr solution (with $x_m$=1:56), we see the NSE coherent data are close to TOF value. This is because the TMA$^+$ does not have long hydro-carbon chains and as a result the internal movement is also lesser than TBA$^+$. Thus both NSE coherent and TOF incoherent data do not differ much for TMA$^+$. But this is not the case for CholineBr (see table \ref{tab:TMA_diff} and \ref{tab:choline_diff}). One alkyl arm of Choline cation is much longer than TMA$^+$ and due to its internal movement, the difference in estimated D$_{tr}$ from incoherent and coherent analysis for Choline$^+$ is higher than for TMA$^+$. This can also be seen from the MSD analysis of the average hydrogen atoms of TMA$^+$, Choline$^+$ and TBA$^+$ cation (figure \ref{TBA-TMA-choline_MSD}) where it is evident that the hydrogen atoms in TBA$^+$ and Choline$^+$ have faster motion below $\sim$400ps due to various internal motion while this is not the case for TMA$^+$.

\subsection{Translational dynamics: anion}    
In this section we briefly present the result (by MD simulation) of Bromide ion (Br$^{-}$) dynamics for the different systems which are studied so far.  

\begin{figure}[!htbp]
\begin{center}
\includegraphics[width=0.45\textwidth,angle=0] {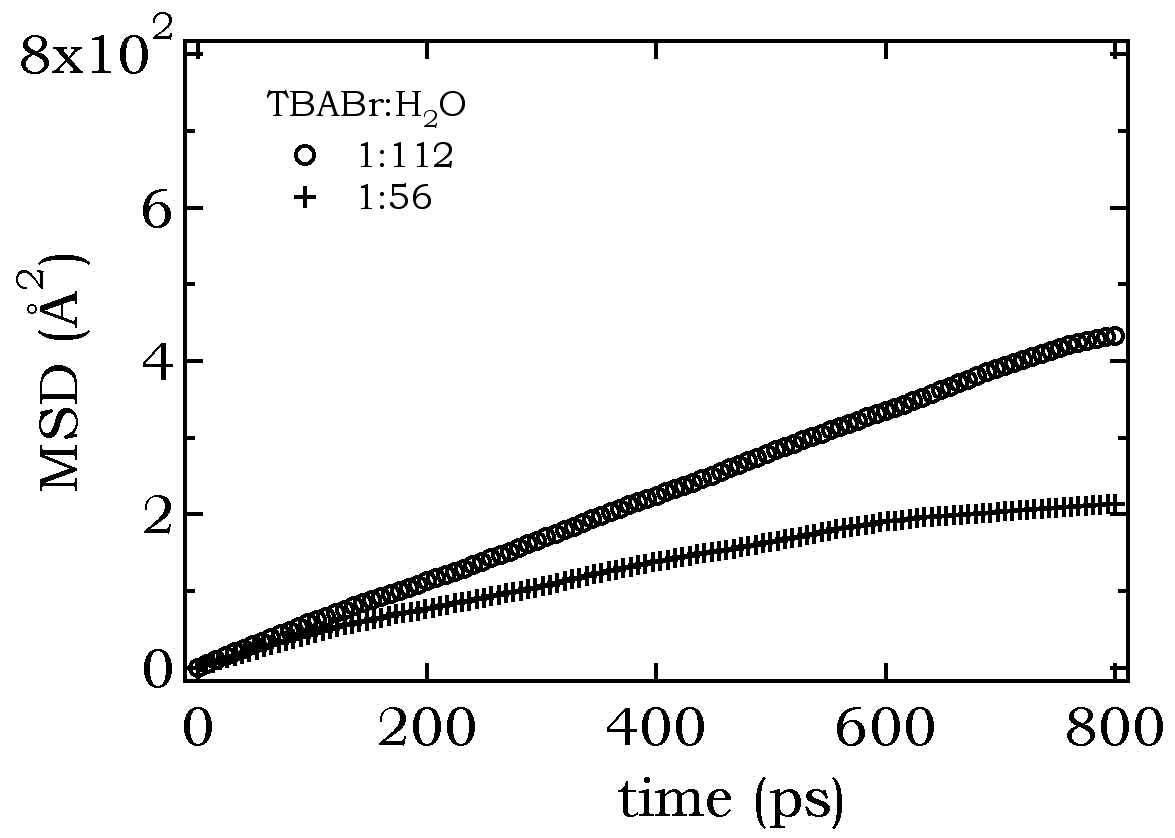}
\vspace{0.1cm}
\bf\caption{The MSD of bromide ions are plotted as a function of time for aqueous solution of TBABr with $x_m$=1:56 and $x_m$=1:112.}
\label{Br_TBABr}
\end{center}
\end{figure}

\begin{table}   
\scriptsize   
\begin{tabular}{|p{1.1cm}|l|p{1.25cm}p{1.25cm}p{1.25cm}|}
	\hline
Concentration & Solution & D$_{tr}$  & of Br$^{-}$ in (10$^{-9}$ m$^2$s$^{-1}$) & \\ 
 	\hline
(salt:H$_2$O) & & Time Scale & Time Scale & Time Scale \\
 & & (0ps to 150ps) & (150ps to 600ps) & (600ps to 800ps) \\
        \hline	        
$x_m$=1:56 & TBABr+H$_2$O & (0.56$\pm$0.01) & (0.39$\pm$0.01) & (0.15$\pm$0.01) \\
        \hline	        
$x_m$=1:112 & TBABr+H$_2$O & (0.74$\pm$0.01) & (0.74$\pm$0.01) & (0.74$\pm$0.01) \\
        \hline
\end{tabular}
 \bf\caption {Extracted translational diffusion coefficient for Br$^-$  extracted by MSD calculation of MD simulation at two different concentration ($x_m$=1:56 and $x_m$=1:112) of aqueous TBABr solution. All the values are corrected by the difference in viscosity between H$_2$O and D$_2$O.}
 \label{tab:Br_diff_1}
\end{table}

\begin{table}   
\scriptsize    
\begin{tabular}{|c|l|cc|}
	\hline
Concentration & Solution & D$_{tr}$ of Br$^{-}$ & in (10$^{-9}$ m$^2$s$^{-1}$) \\ 
 	\hline
(salt:H$_2$O) & & Time Scale & Time Scale \\
 & & (0ps to 400ps) & (400ps to 800ps) \\
        \hline	
$x_m$=1:56 & NaBr+H$_2$O & (0.91$\pm$0.01) & (1.57$\pm$0.01) \\
        \hline	        
$x_m$=1:56 & TMABr+H$_2$O & (0.91$\pm$0.01) & (1.21$\pm$0.01) \\
        \hline
$x_m$=1:56 & ChollineBr+H$_2$O & (1.01$\pm$0.01) & (0.98$\pm$0.01) \\        
       \hline
 & & Time Scale & Time Scale \\
 & & (0ps to 450ps) & (450ps to 800ps) \\
        \hline
$x_m$=1:22 & NaBr+H$_2$O & (0.77$\pm$0.01) & (0.87$\pm$0.01) \\        
       \hline
$x_m$=1:22 & TMABr+H$_2$O & (0.77$\pm$0.01) & (0.92$\pm$0.01) \\        
       \hline
 $x_m$=1:22 & ChollineBr+H$_2$O & (0.66$\pm$0.01) & (0.66$\pm$0.01) \\ 
     \hline	
\end{tabular}
 \bf\caption {Extracted translational diffusion coefficient for Br$^-$ ion in aqueous solution of TMABr, CholineBr and NaBr at two different concentration ($x_m$=1:56 and $x_m$=1:22) extracted by MSD calculation of MD simulation. All the values are corrected by the difference in viscosity between H$_2$O and D$_2$O.}
 \label{tab:Br_diff_2}
\end{table}  

\begin{figure}[!htbp]
\begin{center}
\includegraphics[width=0.45\textwidth,angle=0] {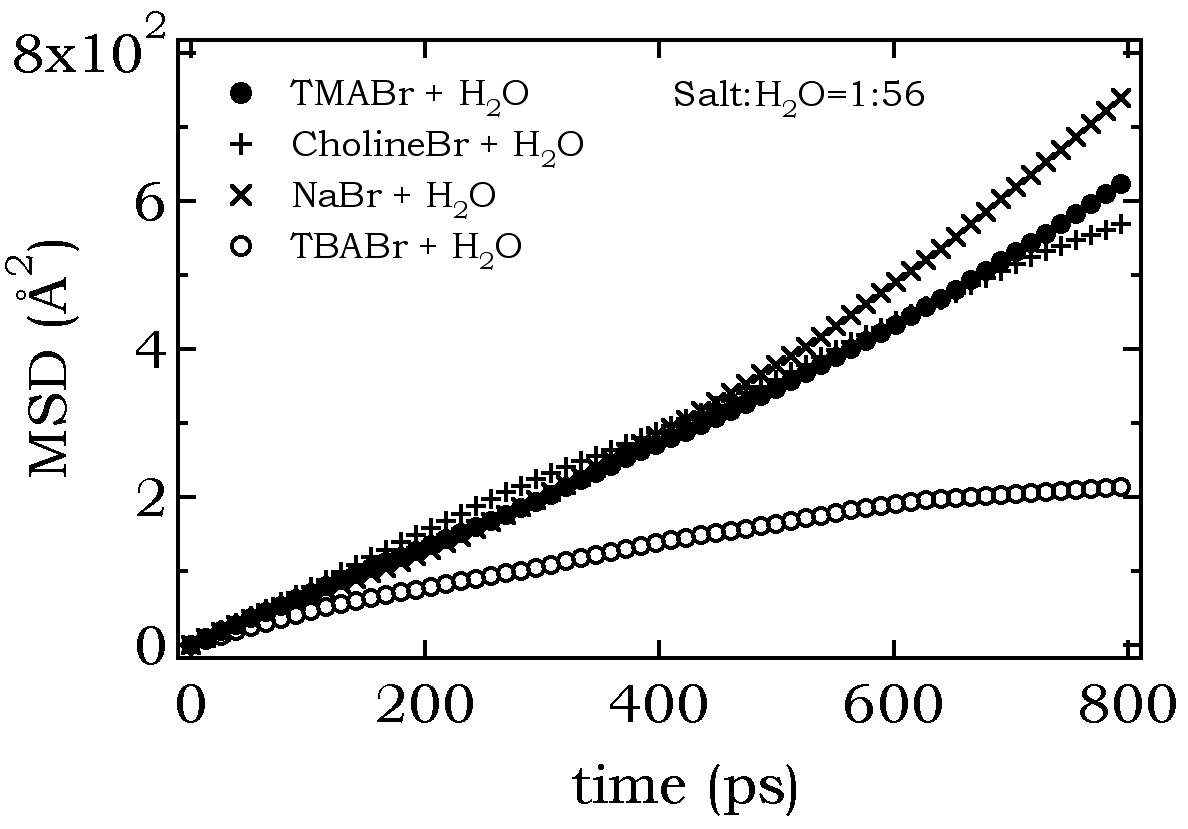}
\includegraphics[width=0.45\textwidth,angle=0] {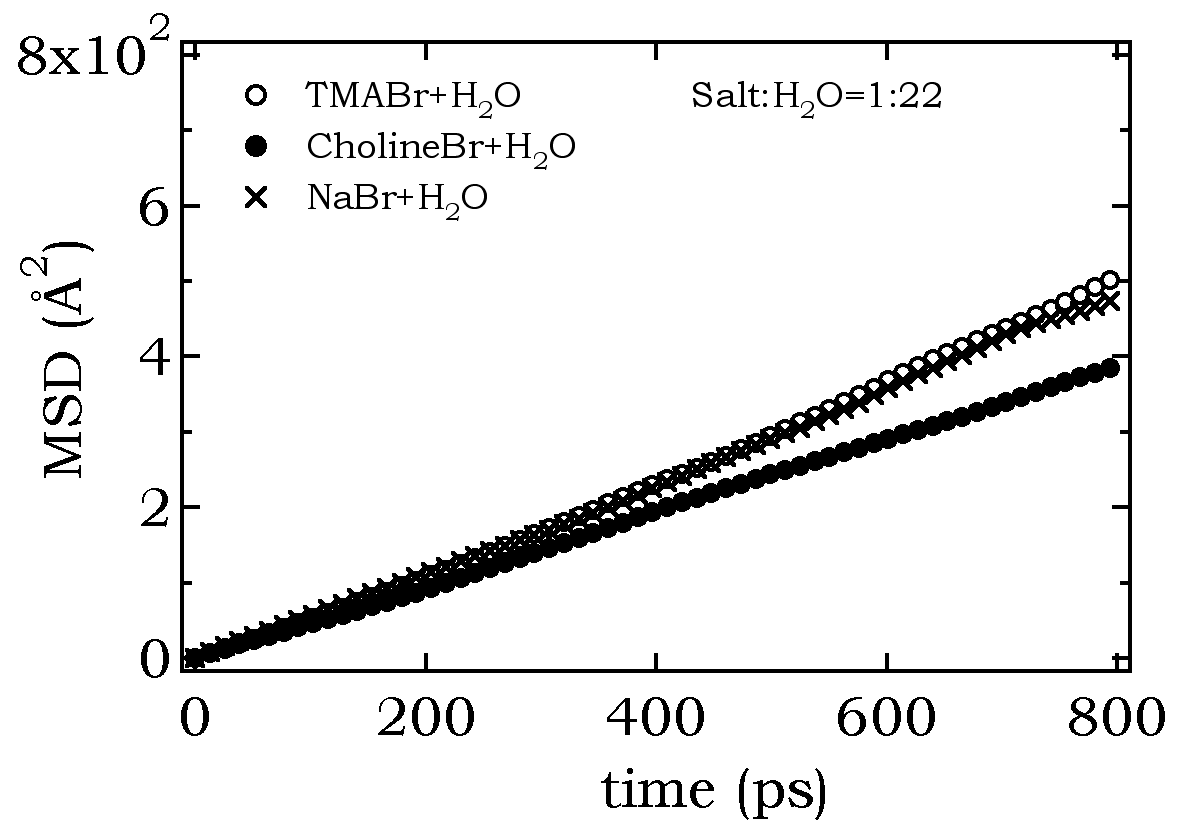}
\vspace{0.1cm}
\bf\caption{The MSD of bromide ions are plotted as a function of time for aqueous solution of NaBr, TMABr and CholineBr with $x_m$=1:56 (top) and $x_m$=1:22 (bottom).}
\label{Br_1-25m}
\end{center}
\end{figure}

The results for aqueous solutions of Na$^+$, TMA$^+$ and Choline$^+$ are summarized in table \ref{tab:Br_diff_2}. We observe that for aqueous solution of $x_m$=1:56, the bromide ion dynamics increases with time for NaBr solution (the effect is less significant for TMABr solution) [figure \ref{Br_1-25m} (top)]. Interestingly this effect is less observed for higher concentration ($x_m$=1:22) [figure \ref{Br_1-25m} (bottom)]. A tentative explanation can be the effect of a medium range order relatively well established when both anion and cation are small and spherical (like Na$^+$, Br$^-$ or even considering TMA$^+$). Then in case of aqueous solution of NaBr or TMABr with $x_m$=1:56, the bromide ion can show an 'excess' of diffusion due to the coulombic forces of ions regularly separated i.e. as the bromide escapes from the electrostatic effects due to the other ions and moves into the inter-ionic space (filled by normal water) its diffusion increases. But as the concentration increases ($x_m$=1:22), the number of normal water molecules in the inter-ionic space decreases and the effect of this excess diffusion is not seen. This kind of behavior is not seen for aqueous CholineBr solution. Because of its longer chain, the bromide ions do not get much free inter-ionic space to show the effect of excess diffusion. 

\section{\label{Conclusion} Conclusion}\   

In this work we have discussed the dynamics and estimated the D$_{tr}$ for TMA and choline for different concentrations. The MD results predict that the Choline cation Centre of Mass (CoM) motion is lower than TMA by a factor of $\sim$1.1 to $\sim$1.3. The difference between D$_{tr}$ extracted from individual hydrogen atom movement and CoM motion (within time window comparable to our TOF measurement) is higher for Choline than for TMA. This is again because of the presence of a longer alkyl chain in the Choline molecule.            

\section{\label{Acknowledgement} Acknowledgement}\ 

The author thanks everyone who helped in this side project during the thesis days. The author notes that there was need to bring the results together from the thesis where some of works were presented in scattered way.    

%This research was sponsored by the U.S. Department of Energy, Office of Science, Office of Basic Energy Sciences, under contract number DE-AC05-00OR22725 with UT-Battelle, LLC. DRC and SG would like to acknowledge support from the US Department of Energy, Office of Basic Energy Sciences, Division of Chemical Sciences, Geosciences and Biosciences, Geosciences Program under grant DE-SC0006878. 
%%%%%%%%%%%%%%%%%%%%%%%%%%
%\clearpage

\section*{\label{Ref} References}\ 

%\end{linenumbers}

\end{document}